\documentclass{PoS}

\usepackage{epsfig}% Include figure files

\PoS{PoS(LAT2005)057}

\rightline{\sffamily UTCCS-P-18, UTHEP-513, KEK-CP-165}\vspace*{-10mm}

\title{Light hadron spectrum and quark masses in 2+1 flavor QCD}

\ShortTitle{Light hadron spectrum and quark masses in 2+1 flavor QCD}

\author{CP-PACS and JLQCD Collaborations:}

\author{\speaker{T.~Ishikawa${}^a$}
        \thanks{E-mail: tomomi@ccs.tsukuba.ac.jp}~,
	S.~Aoki${}^{b,c}$, O.~B\"ar${}^b$, M.~Fukugita${}^d$,
	S.~Hashimoto${}^{e,f}$, K.-I.~Ishikawa${}^g$, N.~Ishizuka${}^{a,b}$,
	Y.~Iwasaki${}^b$, K.~Kanaya${}^b$, T.~Kaneko${}^{e,f}$,
	Y.~Kuramashi${}^{a,b}$, M.~Okawa${}^g$, Y.~Taniguchi${}^{a,b}$,
	N.~Tsutsui${}^e$, A.~Ukawa${}^{a,b}$ and T.~Yoshi\'{e}${}^{a,b}$\\

	${}^a$ Center for Computational Sciences, University of Tsukuba,
 	       Tsukuba, Ibaraki 305-8577, Japan\\
	${}^b$ Graduate School of Pure and Applied Sciences,
	       University of Tsukuba,
	       Tsukuba, Ibaraki 305-8571, Japan\\
	${}^c$ Riken BNL Research Center, Brookhaven National Laboratory,
	       Upton, New York 11973, USA\\
	${}^d$ Institute for Cosmic Ray Research, University of Tokyo,
               Kashiwa 277-8582, Japan\\
        ${}^e$ High Energy Accelerator Research Organization (KEK),
               Tsukuba 305-0801, Japan\\
	${}^f$ School of High Energy Accelerator Science,
  	       The Graduate University for Advanced Studies (Sokendai),
  	       Tsukuba 305-0801, Japan\\
        ${}^g$ Department of Physics, Hiroshima University,
               Higashi-Hiroshima, Hiroshima 739-8526, Japan
	}

\abstract{
CP-PACS and JLQCD collaborations are carrying out a joint
project of the 2+1 flavor full QCD simulation.
Gauge configurations are generated for the non-perturbatively
$O(a)$-improved Wilson quark action
and the Iwasaki gauge action using PHMC algorithm at
three lattice spacings, $a\sim 0.076$, $0.010$ and $0.122$~fm,
with a fixed physical volume $(2.0~\mbox{fm})^3$.
We present analysis for the light meson spectrum 
and quark masses in the continuum limit, which are determined
using data obtained from the simulations at the two coarser lattices. 
Our simulations reproduce experimental values of meson masses. 
The ud and strange quark masses turn out to be 
$m_{ud}^{\overline{MS}}(\mu=2~\mbox{GeV})=3.34(23)~\mbox{MeV}$ and
$m_s^{\overline{MS}}(\mu=2~\mbox{GeV})=86.7(5.9)~\mbox{MeV}$.
We also show preliminary results at our finest lattice spacing
for which simulations are still being continued.
}

\FullConference{XXIIIrd International Symposium on Lattice Field Theory\\
		 25-30 July 2005\\
		 Trinity College, Dublin, Ireland}

\begin{document}

\section{Introduction}

The calculation of the light hadron spectrum and quark masses is 
the most fundamental issue in lattice QCD simulations. 
So far, systematic studies made in quenched~\cite{Aoki:2000-1}
and two flavor ($N_f=2$) full QCD~\cite{Ali_Khan:2000-1} 
revealed that 
1) the $O(10\%)$ deviation of the quenched spectrum from experiment is 
largely reduced in $N_f=2$ QCD, and
2) dynamical up and down quarks reduce significantly the quark masses.
As a next step in this direction, the CP-PACS and JLQCD collaborations
are carrying out a 2+1 flavor ($N_f=2+1$) full QCD simulation
project~\cite{KanekoTIshikawa:2004-1:2005-1},
in which degenerate up and down (``light'') quarks 
and a strange quark are treated dynamically.
With this simulation, we hope to obtain the spectrum and quark masses
much closer to those of the QCD with no approximation.

In this article, we present the results in the continuum limit
extrapolated from two lattice spacings together with preliminary 
results obtained by on-going simulations at our finest lattice spacing.
In this study, we employ the Wilson quark formalism as in our
previous works,
because this formalism has no ambiguity in quark-flavor interpretation. 
We note that a similar attempt is being made~\cite{Aubin:2004-1} 
with the staggered quark formalism.

\section{Gauge configuration generation}

For the lattice action, we employ the renormalization group
(RG) improved Iwasaki gauge action and the clover quark action
with the improvement coefficient $c_{SW}$ determined 
non-perturbatively for the RG action~\cite{Aoki:2005-1}.
The choice of the gauge action is made to avoid a first-order phase
transition (lattice artifact) observed for the plaquette gauge 
action \cite{Aoki:2004-bulk}.

Configurations are generated with the Polynomial Hybrid Monte Carlo
(PHMC) algorithm.
(See Ref.\cite{Aoki:2002-2} for our implementation.) 
The time step $\delta\tau$ in the molecular dynamics
and the order $N_{poly}$ of Chebyshev polynomial used for
an approximation of quark determinant are chosen such that 
the HMC and the global Metropolis acceptance rate achieves 
$85\%$ and $90\%$, respectively.

Simulations are performed at three values of the coupling constant
chosen so that the square of the lattice spacing $a^2$ is placed
at an even interval.  
The physical volume is fixed at $(2.0~\mbox{fm})^3$. 
Simulation parameters are listed in 
Table~\ref{TAB:simulation_parameters}.
At each coupling, we generate configurations for ten combinations
of hopping parameters $(\kappa_{ud},\kappa_s)$, five for 
the ud quark mass taken in the range $m_{PS}/m_V\sim 0.6-0.78$ 
and two for the strange quark mass chosen around $m_{PS}/m_V\sim 0.7$.
Gauge configuration generation has already been finished at
the two coarser lattices. We plan to continue the on-going simulation
at the finest lattice up to $6000$ trajectories.

\begin{table}
\begin{center}
\begin{tabular}{ccccc}
 \hline
 $\beta$ & size & $a$ [fm] ($K$-input) & $a$ [fm] ($\phi$-input) &
 trajectory \\
 \hline
 $1.83$ & $16^3\times32$ & $0.1222(17)$ & $0.1233(30)$ & $7000 - 8600$ \\
 $1.90$ & $20^3\times40$ & $0.0993(19)$ & $0.0995(19)$ & $5000 - 9200$ \\
 $2.05$ & $28^3\times56$ & $0.0758(48)$ & $0.0755(48)$ & $3000 - 4000$ \\
 \hline
\end{tabular}
\caption{Simulation parameters.
         The production run at $\beta=2.05$ is still in progress.}
\label{TAB:simulation_parameters}
\end{center}
\end{table}

\section{Measurement and analysis}

Measurements are performed at every 10 trajectories.
We first fix the configuration to the Coulomb gauge and
then calculate quark propagators for valence quark masses 
taken equal to one of sea quark masses, 
using point and exponentially smeared sources and sinks. 
For analysis, we use the combination of smeared source and point sink, 
because with this combination effective masses reach a plateau
earliest and the statistical error is the smallest.
Meson masses and quark masses are determined from single mass 
$\chi^2$ fits to correlators $\langle P(t)P(0)\rangle$, 
$\langle V(t)V(0)\rangle$ and $\langle A_4(t)P(0)\rangle$, 
where $P$, $V$ and $A_{\mu}$ denote the pseudoscalar, 
the vector and the one-loop $O(a)$-improved axial-vector current, 
respectively. 
We include correlations in time but ignore correlation among correlators.
Errors are estimated by the binned jackknife method with a bin size of
$100$ trajectories. 
 
Chiral fits are made to light-light, light-strange and
strange-strange meson masses simultaneously ignoring their correlations,
using a quadratic polynomial function in terms of the sea and 
valence quark masses.
\footnote{Chiral fit is one of the important sources of systematic errors.
	  Recently, Wilson chiral perturbation theory (WChPT) has been
          proposed, in which infrared chiral
	  logarithms and finite lattice spacing corrections to them for
	  the Wilson quark action are incorporated.
	  An application of WChPT to our $N_f=2+1$ QCD data
          is in progress. (See Ref.~\cite{Aoki:2003-WChPT_Nf2}.)}
For chiral fits, we use two definitions of the quark mass. 
One is the vector Ward identity (VWI) quark mass defined by 
$m_q^{VWI}=(1/\kappa-1/\kappa_c)/2$, where $\kappa_c$ is the
critical hopping parameter at which $m_{PS}$ at 
$\kappa_{ud}=\kappa_s=\kappa_{val}=\kappa_c$ vanishes.
The other is the axial-vector Ward identity (AWI) quark mass 
defined by $m_q^{AWI}=\nabla_{\mu}A_{\mu}(x)/(2P(x))$.
The lattice spacing, the meson masses and $m_q^{VWI}$ at the physical
point are determined from the chiral fit in terms of $m_q^{VWI}$, while
$m_q^{AWI}$ at the physical point are from the chiral fit in terms of 
$m_q^{AWI}$.
The lattice spacing and the meson masses obtained
from the latter chiral fit are consistent with those from the former.

For the physical point, we consider two cases. One is called
as ``$K$-input'' in which we take the experimental values  
$m_{\pi}=0.1350~\mbox{GeV}$, $m_{\rho}=0.7684~\mbox{GeV}$
and $m_K=0.4977~\mbox{GeV}$ as inputs.
The other is ``$\phi$-input'' in which $m_{\pi}$, $m_{\rho}$ and
$m_{\phi}=1.0194~\mbox{GeV}$ are taken as inputs.
The lattice spacings determined from
the $K$- and $\phi$- inputs are consistent with each other,
as shown in Table~\ref{TAB:simulation_parameters}.

\section{Light meson spectrum}

\begin{figure}
\begin{center}
\includegraphics[scale=0.56, viewport = 0 0 750 330, clip]
 {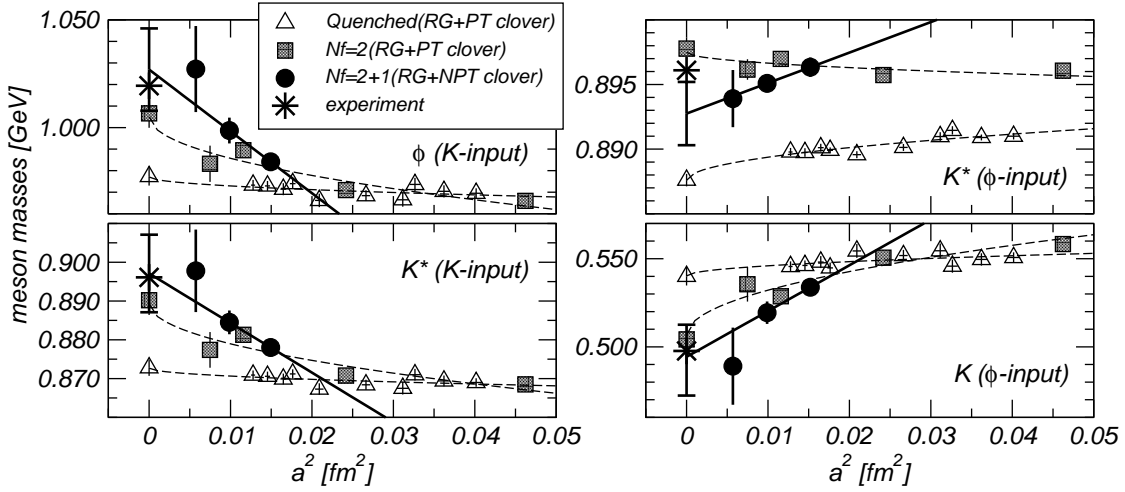}
\caption{Continuum extrapolation of meson masses,
         compared with those for quenched and $N_f=2$ 
         QCD~\protect\cite{Ali_Khan:2000-1}.
         Note that the quenched and $N_f=2$ simulations are made with 
         the one-loop perturbatively $O(a)$-improved clover action.
         Thus extrapolations are made linearly in $a$.}
\label{FIG:meson_masses}
\end{center}
\end{figure}

We evaluate the meson masses in the continuum limit extrapolating the
data linearly in $a^2$ at two coarser lattices,
for which runs and measurements are completed.
The results shown in Fig.~\ref{FIG:meson_masses} are consistent with
experiment though errors are still large. 
Taking $K^*$ with $K$-input as an example, the linear fit is given by 
$m_{K^*}=m_0(1+c(\Lambda_{QCD}\cdot a)^2)$ with
$m_0=897(10)$~MeV and $c=-1.38(78)$ where we assumed
$\Lambda_{QCD}=200$~MeV.
The O(1) magnitude of the coefficient $c$ is reasonable,
albeit apparently larger scaling violation is somewhat disappointing,
in comparison to quenched(triangles) and $N_f=2$(squares)
data~\cite{Ali_Khan:2000-1} which are based on only the tadpole-improved
action.   
The points at the finest lattice spacing are based on measurements for
$3000$ trajectories.
Statistics are still low.
We hope we continue simulations up to the full $6000$ trajectories and
clarify the continuum extrapolation in the near future. 

\section{Light quark masses}

The physical quark mass is determined for the $\overline{\mbox{MS}}$ 
scheme at the scale $\mu=2$~GeV. To do this, we first translate
the quark mass to that in the $\overline{\mbox{MS}}$ scheme at scale 
$\mu=a^{-1}$ using $Z$-factors determined by tadpole-improved
one-loop perturbation theory~\cite{Aoki:1998-PT-renorm}.
The renormalized quark masses are then evolved to $\mu=2$~GeV
using the four-loop RG-equation.

\begin{figure}
\begin{center}
\includegraphics[scale=0.43, viewport = 0 0 550 325, clip]
 {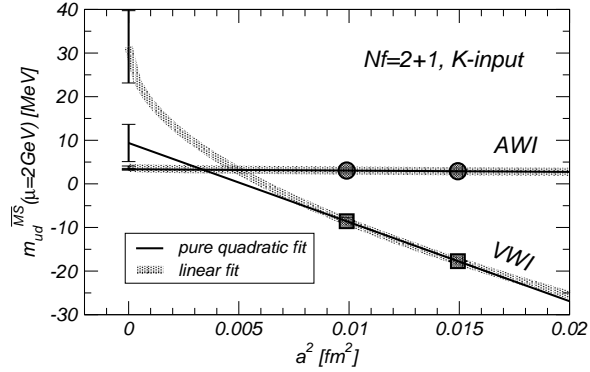}
\caption{Continuum extrapolations of the up and down quark mass 
         defined by the AWI and VWI ($K$-input).}
\label{FIG:AWI_and_VWI_ud}
\end{center}
\end{figure}

In principle, the quark mass calculated in our way can have $O(g^4a)$
scaling violation since we use the matching factor and the improvement
coefficient $c_A$ of the axial-vector current determined by one-loop
perturbation theory.
We compare in Fig.~\ref{FIG:AWI_and_VWI_ud} the ud quark masses
calculated with the VWI and AWI definitions.
They do not agree at all in the continuum limit,
when they are extrapolated in $a$.
On the other hand, pure quadratic fits yield reasonably consistent values.
This suggests that terms linear in $a$ are small for ud quark masses.
Similarly, the pure quadratic extrapolations of the VWI and AWI strange
quark masses give a common value in the continuum limit.

We take the AWI definition to quote the continuum limit for estimates of 
the quark mass, since the magnitude of scaling violation is smaller for
$m_q^{AWI}$ than for $m_q^{VWI}$, and thus the error
in the continuum limit is smaller for $m_q^{AWI}$, as shown in 
Fig.~\ref{FIG:AWI_and_VWI_ud}.\footnote{
The VWI quark mass for the ud quarks is negative at our simulation 
points. This originates from a lack of chiral symmetry of the Wilson
quark action. This is another reason to prefer the AWI definition.}
Note that this property is also observed in $N_f=2$ QCD. 

\begin{figure}
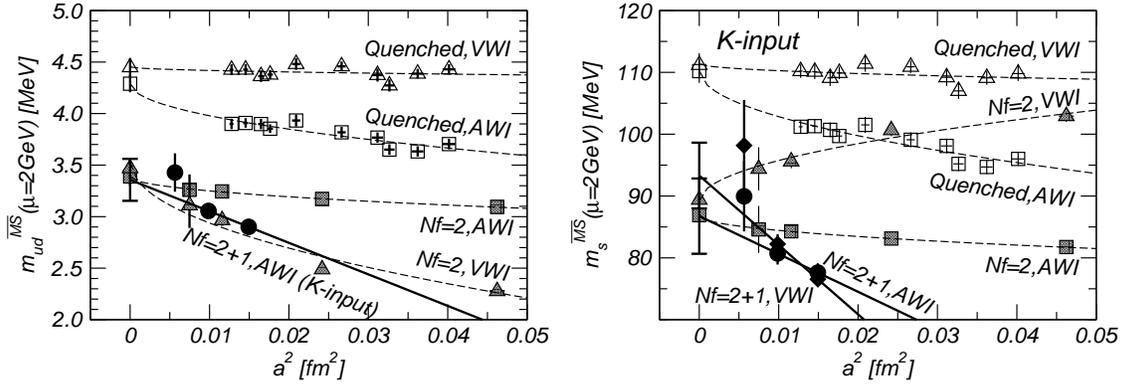

\begin{center}
\begin{minipage}{7.4cm}
\includegraphics[scale=0.49, viewport = 0 0 430 295, clip]
 {Mud_vs_a2_k-input.eps}
\end{minipage}
\begin{minipage}{7.4cm}
\includegraphics[scale=0.49, viewport = 0 0 430 295, clip]
 {Ms_vs_a2_k-input.eps}
\end{minipage}
\caption{Continuum extrapolations of the up, down and strange quark masses
         obtained with the $K$-input.
         The data at the finest lattice is not included in
         the continuum extrapolations.
         For comparison, results for quenched and $N_f=2$ QCD are
         overlaid.}
\label{FIG:uds_quark_masses}
\end{center}
\end{figure}

In Fig.~\ref{FIG:uds_quark_masses},
we present continuum extrapolations of the up, down and strange quark
masses determined with the $K$-input. The ud quark mass with
the $\phi$-input agrees with the $K$-input at each lattice spacing.
For the strange quark mass, the $K$- and the $\phi$-inputs give
different values at finite lattice spacings,
but masses are extrapolated to a common value in the continuum limit,
as observed in $N_f=2$ QCD.
Thus we estimate the quark masses in the continuum limit from 
a combined fit to data with the $K$- and the $\phi$-inputs,
\begin{equation}
 m_{ud}^{\overline{MS}}(\mu=2~\mbox{GeV})=3.34(23)~\mbox{MeV},\;\;\;\;\;
 m_s^{\overline{MS}}(\mu=2~\mbox{GeV})=86.7(5.9)~\mbox{MeV}.
\end{equation}
In the continuum limit, the quark masses estimated for the $N_f=2+1$
QCD do not show deviations beyond statistical errors from those of
$N_f=2$ QCD.

\section{Conclusions and future plans}

In this article we have reported the status of our Wilson-clover $N_f=2+1$
simulations.  
The spectrum at two lattice spacings, when extrapolated to
the continuum limit are consistent with experiment.
However, the outcome of the on-going runs at the finest lattice spacing
at $a\sim 0.076$~fm, which we hope to complete in a half a year or so,
is needed for a precise conclusion.
The same comment also applies to light quark masses.
A shortcoming of the current runs is a relatively large value of
the dynamical up and down quark masses.  
We hope to bring improvement in our program by a combined 
application of the improved algorithm provided by the domain
decomposition idea~\cite{Luscher:2003} and an enhanced computing power
to be provided by PACS-CS, the successor to the CP-PACS computer. 

\begin{acknowledgments}
This work is supported by 
the Epoch Making Simulation Projects 
of Earth Simulator Center,
the Large Scale Simulation Program No. 132 (FY2005) of
High Energy Accelerator Research Organization (KEK),
the Large Scale Simulation Projects of
Academic Computing and Communications Center, University of Tsukuba,
Super Sinet Projects 
of National Institute of Infomatics,
and also by the Grant-in-Aid of the Ministry of Education
(Nos. 13135204, 13640260, 14046202, 14740173, 15204015, 15540251,
      15740134, 16028201, 16540228, 16740147, 17340066, 17540259).
\end{acknowledgments}
\vspace*{-5mm}

%%%%%%%%%%%%%%%%%%%%%%%%%%%%%%%%%%%%%%%%%%%%%%%%%%%%%%%%%%%
%%%%%%%%%%%%%%%%%%%%%% References %%%%%%%%%%%%%%%%%%%%%%%%%
%%%%%%%%%%%%%%%%%%%%%%%%%%%%%%%%%%%%%%%%%%%%%%%%%%%%%%%%%%%

\end{document}